\documentclass[reprint,amsmath,amssymb,aps,prl,superscriptaddress]{revtex4-1}
\usepackage{graphicx}
\usepackage{dcolumn}
\usepackage{bm}
\usepackage{float}
\usepackage{amsmath}
\usepackage{color}
\begin{document}
\setlength{\parskip}{0pt}

\title{Active suspensions of bacteria and passive objects: a model for the near field pair dynamics}
\author{Bokai Zhang}
\affiliation{Beijing Computational Science Research Center, Beijing 100193, China}
\affiliation{Department of Physics, Zhejiang Sci-Tech University, Hangzhou 310018, China}
\author{Yang Ding}
\email{dingyang@csrc.ac.cn}
\affiliation{Beijing Computational Science Research Center, Beijing 100193, China}
\author{Xinliang Xu}
\email{xinliang@csrc.ac.cn}
\affiliation{Beijing Computational Science Research Center, Beijing 100193, China}
\affiliation{Department of Physics, Beijing Normal University, Beijing 100875, China}

\begin{abstract}

Near field hydrodynamic interactions are essential to determine many important emergent behaviors observed in active suspensions, but have not been successfully modeled so far.  In this work we propose an effective model capable of efficiently capturing the essence of the near field hydrodynamic interactions, validated numerically by a pedagogic model system consisting of an \emph{E. coli} and a spherical tracer.  The proposed model effectively captures all the details of near field hydrodynamics through only a tensorial coefficient of resistance, which is fundamentally different from, and thus cannot be replaced by, an effective interaction of conservative nature.  In a critical test case that studies the scattering angle of the bacterium-tracer pair dynamics, calculations based on the proposed model reveals a region in parameter space where the bacterium is trapped by the spherical tracer, a phenomenon that is regularly observed in experiments but cannot be explained by any existing model.

\end{abstract}

\maketitle

Dense suspensions of microorganisms swimming in complex environments are ubiquitous in nature.  The hydrodynamic interactions among the micro-swimmers and the surrounding boundary give rise to many interesting phenomena.  For instance, the motions of micro-swimmers perturb the otherwise quiescent fluid resulting in flow fields that give rise to enhanced diffusion of other suspended objects \cite{RN1,RN2,RN3,RN4,RN5}, which can be related to many biologically important processes such as the nutrients transportation \cite{RN6} and biomixing \cite{RN7,RN8}.  At the same time, the dynamical behavior of each micro-swimmer is also strongly affected by the ambient objects, leading to intriguing collective dynamics \cite{RN9} and macroscopic behaviors \cite{RN10} that inspire innovative microfluidic devices \cite{RN11}.  While the near field hydrodynamic interactions have been shown to be essential to many collective behaviors \cite{RN12,RN13,RN14,RN15,RN16} (e.g. the emergence of flocking, the so-called active turbulence induced by the hydrodynamic instability), a minimal model that correctly capture the essence of the near field hydrodynamics for general micro-swimmers is not available yet and is the subject of this paper.
\begin{figure}[b]
\includegraphics[width=.35\textwidth]{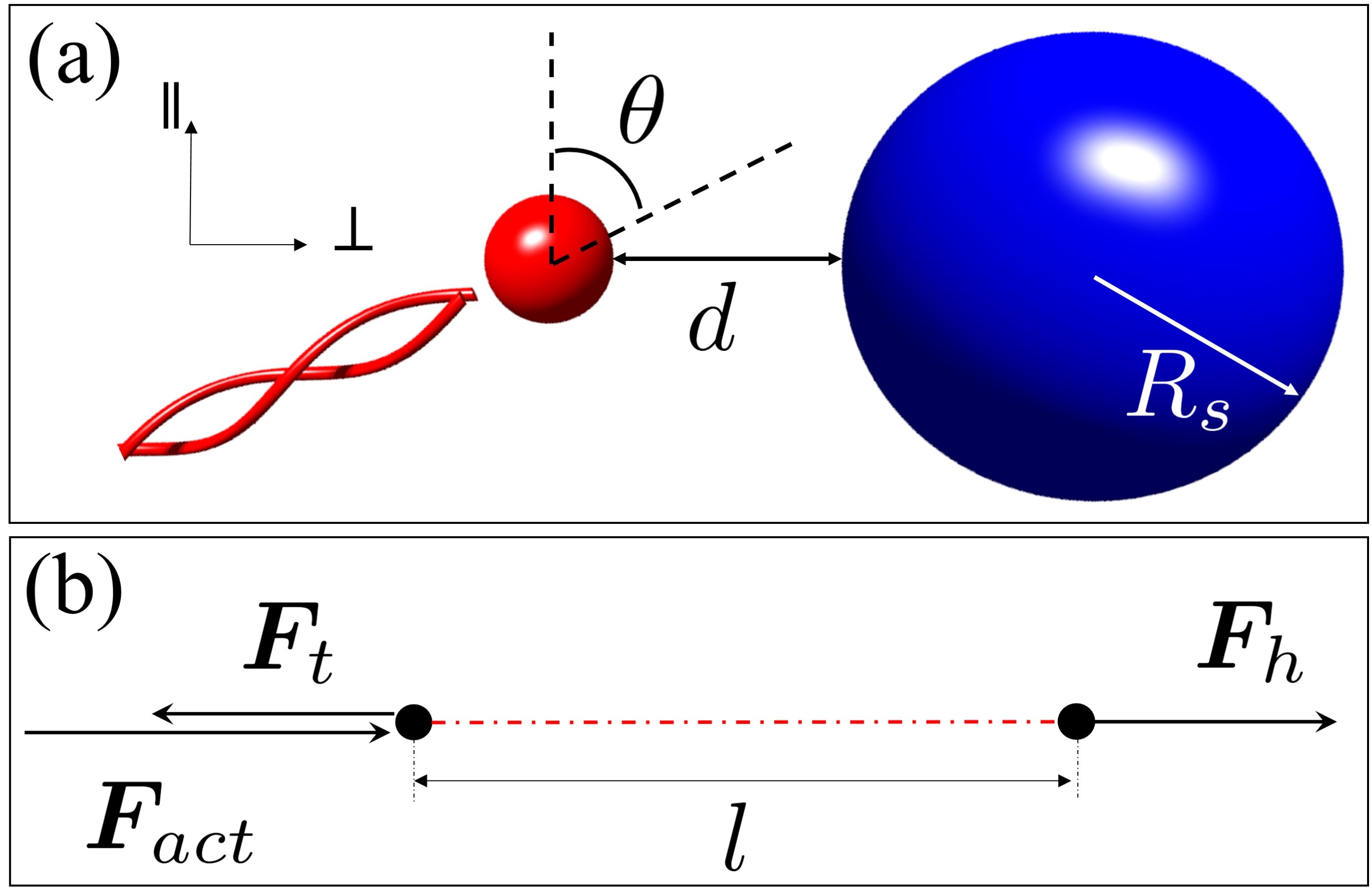}
\caption{The system at study. (a) The system consists of one bacterium (red) and one tracer (blue). (b) In our proposed model, the bacterium is described as two point-forces ($\bm{F}_h=\bm{F}_{eff}$ for the head and $\bm{F}_t=-\bm{F}_{eff}$ for the tails) separated by a distance $l$. The propulsive force arising from the spinning of the helical tails is $\bm{F}_{act}$.}
\end{figure}

Given its characteristic small size, each free-swimming bacterium can be considered as force-free and torque-free, thus at the lowest order a force dipole \cite{RN17, RN18}.  This simple dipole model has been quite successful in predicting the qualitative behaviors of quantities such as the enhanced diffusion of passive tracer particles \cite{RN1}.  But the model prediction is only accurate for the far field where the size of the bacterium is negligibly small, and requires a cutoff size to avoid the unphysical divergence at the location of the force dipole.  In a more realistic model where one bacterium is simplified as a rigid dumbbell of two beads connected by a rigid rod, the characteristic size of the bacterium is incorporated by treating the bacterium as two point-forces separated by a length of $l$.  In this two-bead model, the tail bead is propelled by a force $\bm{F}_{act}$, provided by flagella that are not treated explicitly.  Then the force balance for the two beads gives that the forces exerted on fluid are simply $\bm{F}_{eff}$ and $-\bm{F}_{eff}$ for the head and tail bead, respectively.  Models of this type have been used to study a variety of systems, including the motion of one bacterium near a plane wall \cite{RN19}, two hydrodynamically interacting bacteria \cite{RN4,RN20}, and the collective motions in a suspension \cite{RN21, RN22}.  However, the treatment of the head and tail beads as two point-forces is unable to capture the characteristics of force distributions essential for the near-field hydrodynamic interaction, leading to some severe unphysical consequences, e.g., invariant $\bm{F}_{eff}$ and hence invariant $\bm{v}_{bac}$ (bacterial self-swimming velocity with respect to background flow) independent of the surrounding environment \cite{RN21, RN22} that inevitably leads to artificial overlaps between bacteria. To solve the catastrophic overlap problem shared by this type of models, one brute force approach commonly used is to introduce a repulsive steric pair interaction.  But the use of an effective repulsion in place of the true near field hydrodynamics may result in problems such as incorrect estimate of $\bm{F}_{eff}$ and unrealistically fast separation for bacterial pairs in close proximity.  The near field hydrodynamic interaction between a pair of swimmers has been studied more carefully in the so-called squirmer model where each bacterium is modelled as one sphere with a prescribed tangential velocity on its surface \cite{RN23}.  Yet it is not clear if the squirmer model is generally applicable to bacteria of arbitrary non-spherical shapes.

In this work, we examine the key ingredients of the existing two-bead model upon which we can build our model, as well as the defects of the existing two-bead model that need to be corrected. To do so, we first numerically study the hydrodynamic interaction between one passive sphere and one bacterium in detail.  The system at study consists one spherical tracer with radius $R_s$ that is force-free and torque-free, and one free-swimming \emph{E. coli} shaped bacterium with fixed bacterium motor rotation rate $\omega_0$, both immersed in a fluid of viscosity $\mu$ (Fig. 1a).  The bacterium is modeled as an assembly of a spherical head of diameter $\sigma$ that is used as the unit for length in our study, and two helical flagella.  The hydrodynamic interactions of the system are investigated at different configurations defined by the surface distance $d$, incoming angle $\theta$ (we have used the same definition as earlier studies \cite{RN24}, where negative $\theta$ corresponds to the ¡°nose down¡± situation with bacterium moving towards the tracer), and the tracer sphere radius $R_s$, as illustrated in Fig. 1a.  As the characteristic size and speed of most bacteria are about 1 $\mu$m and 1 $\mu$m/s, respectively, in water the corresponding Reynolds number is very low ($10^{-5}\sim 10^{-2}$).  Then the surrounding fluid can be well described by the so-called creeping flow dictated by the linear Stokes equation.  And the hydrodynamic interaction can be quantitatively evaluated by solving the linear Stokes equation, with no-slip boundary condition on the surfaces of the sphere and the bacterium.  The Stokes equation with moving boundaries can be routinely solved using the numerical method of Stokeslets \cite{RN25}, where the boundary surfaces are divided into a large number of small regions and the force distributed on each region is then approximated by a point force.  This method is based on the fact that the creeping flow $\bm{u}$ at location $\bm{r}'$ due to each point force $\bm{f}$ at location $\bm{r}_0$ is analytically available as
\begin{equation}
\bm{u}(\bm{r}')=\bm{G}(\bm{r}',\bm{r}_0)\bm{f}(\bm{r}_0),
\end{equation}
where the Green¡¯s function $\bm{G}$ is a fundamental solution to the linear Stokes equation and is called a Stokeslet.  The Stokeslet in three dimensions manifests in the tensor form of $G_{ij} (\bm{r}',\bm{r}_0 )=\frac{1}{8 \pi \mu} (\frac{\delta_{ij}}{r}+\frac{r_i r_j}{r^3})$, with $r\equiv|\bm{r}|\equiv|\bm{r}'-\bm{r}_0 |$.  Then the solution of the entire flow field is the sum of all the flows each generated by one of these point forces.  For more information about our pedagogic system and the method of Stokeslets, please see Supplemental Information (SI).

The dependence of instantaneous speed of the spherical tracer, $|\bm{v}_s|$, on the surface distance $d$ at a few typical incoming angles $\theta$ and sphere radii $R_s$ are shown in Fig. 2.   At large $d$, $|\bm{v}_s|$ always decays as $d^{-2}$, regardless of $\theta$ and $R_s$.  This power-law decay is consistent with previous experimental observations \cite{RN18}, as well as the predictions of the dipole model \cite{RN17}.  At intermediate $d$, $|\bm{v}_s|$ behaves qualitatively different from the dipole predictions.  And for large tracer sphere cases, e.g. $R_s=50$, a nonmonotonic behavior of $|\bm{v}_s|$ is observed.

To explain the behavior of tracer motion at intermediate $d$, we follow the idea of existing two-bead model that treats the bacterium as two point-forces.  To do so we add all the point forces on bacterium head as obtained in our Stokeslets method and place the sum $\bm{F}_{eff}=\sum_{n=1}^{N_{head}}\bm{f}_n$  at the geometric center of the head, and add all the point forces on bacterium tails and place the sum ($-\bm{F}_{eff}$, as dictated by the force-free condition) at the geometric center of the tails.  Under the influence of these two point-forces that are separated by a distance of $l$ as illustrated in Fig. 1b, the motion of the tracer sphere follows the Faxen's law:
\begin{equation}\label{Eq2}
  \bm{v}_s=\Big (1+\frac{R_s^2}{6}\nabla^2\Big )\bm{u}(\bm{r})\Big |_{\bm{r}=\bm{R}_0},
\end{equation}
where $\bm{R}_0$ is the location of the center of the sphere, $\bm{u}(\bm{r})$ is the flow field generated by the two point-forces (the flow field generated by each point-force can be obtained through Eq. 1).  As illustrated in Fig. 2, the predictions of $\bm{v}_s$ obtained from Eq. 2 agree very well with our numerically obtained results from the Stokeslets method at both large and intermediate $d$.  The nonmonotonic behavior observed for large tracer cases is interpreted as the result of a competition between the sphere size and the characteristic length defined by the gradient of the flow field that is dictated by $l$.
\begin{figure}[b]
\includegraphics[width=.35\textwidth]{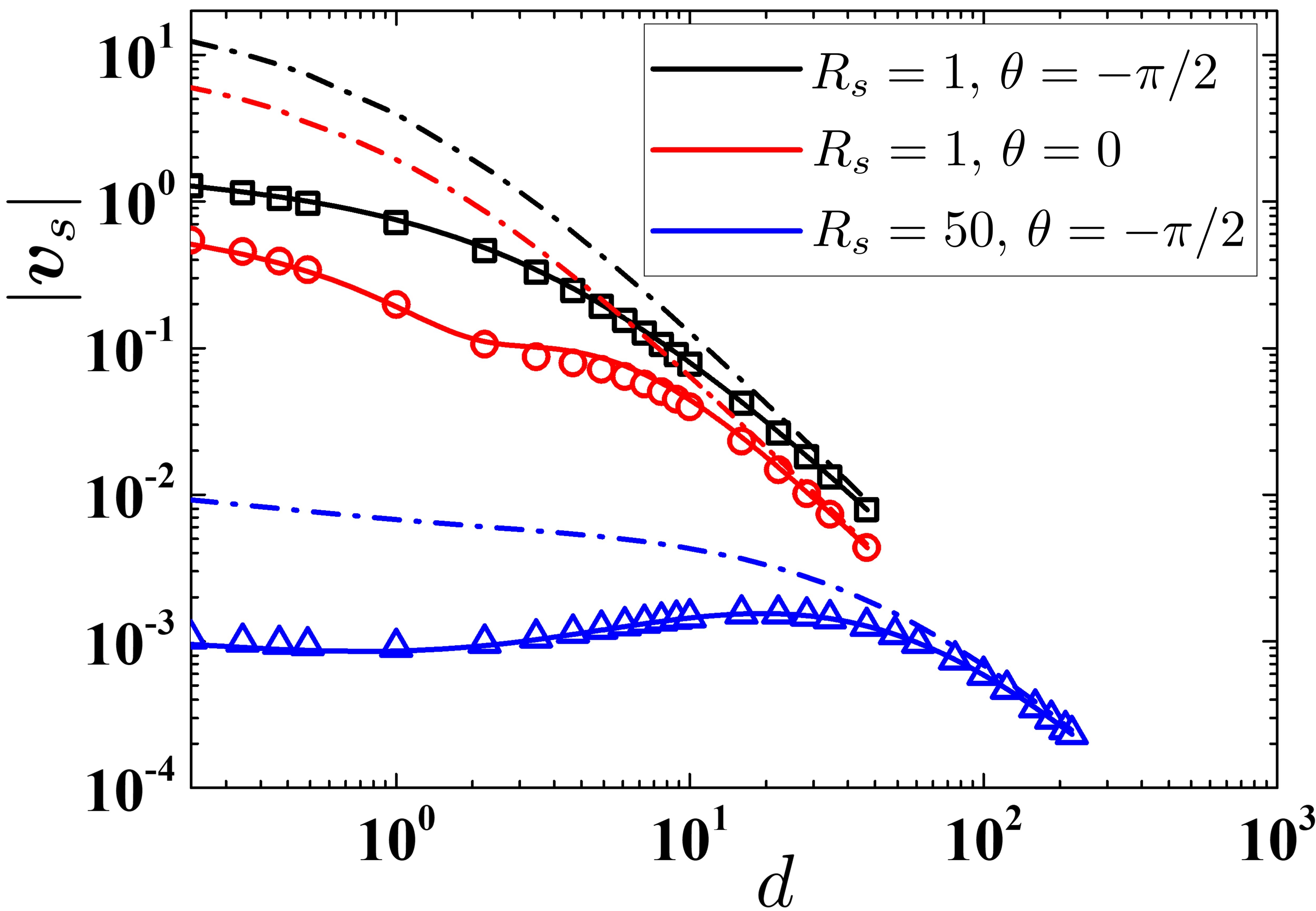}
\caption{The motion of the spherical tracer, $|\bm{v}_s|$, as a function of $d$ for a few typical $\theta$ and $R_s$. The open symbols are the results from the method of Stokeslets. The dashed lines and solid lines are model predictions where the bacterium is modeled as a force dipole, and two point-forces with fixed separation $l$, respectively.}
\end{figure}

The above quantitative agreement regarding $\bm{v}_s$ is achieved by placing the two point-forces at fixed locations (one at the geometric center of the head and the other at the geometric center of the tails) independent of the configurational parameters \{$d$, $\theta$, $R_s$\}, which supports a key ingredient of the existing two-bead model that treat the locations of the two point-forces as bacterial intrinsic properties.  By evaluations of bacteria with other shapes of tails, we further demonstrated that the locations of the two point-forces are indeed bacterial intrinsic properties (Fig. S5 in SI).

However, unlike the predictions of the existing two-bead model that bacterial motion $\bm{v}_{bac}$ and the force $\bm{F}_{eff}$ are both invariant, our numerical results show otherwise.  This qualitative difference can be best illustrated as in Fig. 3a, where we show $|\bm{v}_{bac} |$ and $|\bm{F}_{eff}|$ (inset) as functions of $d$, at $\theta=-\pi/2$ (the bacterium moving towards the center of the spherical tracer) for two typical sphere radii $R_s=1$ and $R_s=100$, respectively.  For $d\gg 1$, the influence of the tracer sphere on the bacterium is negligible so that $|\bm{v}_{bac}|$ and $|\bm{F}_{eff}|$ reduce to their corresponding values for bacterium swimming in free space: $v_{bac}^0$ and $F_{eff}^0$, respectively. At smaller $d$, we see a significant decrease in bacterium swimming velocity, from $v_{bac}^0$ to $|\bm{v}_{bac} |(d=0.1)\approx 0.3v_{bac}^0$ in the system with $R_s=100$, presumably due to the increase of effective resistance felt by the bacterium.  In the same small $d$ regime, we see that $|\bm{F}_{eff}|$ increases noticeably from $F_{eff}^0$.

We propose that the key to the observed strong dependence of $\bm{v}_{bac}$ and $\bm{F}_{eff}$ on surface distance $d$ is the near field hydrodynamic interaction between the bacterium and the sphere, which can be quantitatively modeled by the resistance tensor $\bm{\xi}$, defined as in:
\begin{equation}\label{Eq3}
  \left(
    \begin{array}{c}
      \bm{F}_h \\
      \bm{F}_s \\
      \bm{F}_t \\
    \end{array}
  \right) = \bm{\xi}\cdot
  \left(
    \begin{array}{c}
      \bm{v}_h \\
      \bm{v}_s \\
      \bm{v}_t-\bm{v}_0 \\
    \end{array}
  \right){\rm with}\ \bm{\xi}=
  \left(
    \begin{array}{ccc}
      \bm{\xi}_{hh} & \bm{\xi}_{hs} & \bm{\xi}_{ht} \\
      \bm{\xi}_{sh} & \bm{\xi}_{ss} & \bm{\xi}_{st} \\
      \bm{\xi}_{th} & \bm{\xi}_{ts} & \bm{\xi}_{tt} \\
    \end{array}
  \right),
\end{equation}
where $\bm{F}_h=\bm{F}_{eff}$, $\bm{F}_t=-\bm{F}_{eff}$, and $\bm{F}_s$ are the forces exerted by the bacterial head, tails, and spherical tracer, respectively (Fig. 1b); $\bm{v}_h$, $\bm{v}_t$, and $\bm{v}_s$ are the velocities of the bacterial head, tails, and spherical tracer, respectively; and $\bm{v}_0\equiv\bm{\xi}_{tt}^{-1}\cdot\bm{F}_{act}$.  Since the tail flagella are very thin ($\sim 10$ nm) comparing to the head ($\sim 1$ $\mu$m), in the simplest consideration we can assume that the tails are not as affected by the near field hydrodynamic interactions.  That is, the force $\bm{F}_{act}$ arising from the spinning of the asymmetrical tails around the longitudinal direction are considered as configuration independent, and the terms in $\bm{\xi}$ involving the tails remain their far field values.  Then the near field hydrodynamic interaction only appears in tensor elements $\bm{\xi}_{hh}$, $\bm{\xi}_{hs}$, and $\bm{\xi}_{ss}$; and we can solve $\bm{F}_{eff}$, and $\bm{v}_{bac}$ as functions of $\bm{F}_{act}$ and $\bm{\xi}$.  In the $d\to 0$ limit, lubrication theory shows that $\bm{\xi}_{hh}$, $\bm{\xi}_{hs}$, and $\bm{\xi}_{ss}$ can all be written as analytic functions of only one parameter, the non-dimensional surface distance $\frac{2d}{0.5+R_s}$ \cite{RN26}.  To keep our model simple, at finite $d$ we write $\bm{\xi}_{hh}$, $\bm{\xi}_{hs}$, and $\bm{\xi}_{ss}$ by extrapolating the analytical lubrication forms (so that near field hydrodynamic interactions are described only by the single parameter $\frac{2d}{0.5+R_s}$), which we use to numerically solve for $\bm{F}_{eff}$, and $\bm{v}_{bac}$ (details of the solution process available in SI).

As illustrated in Fig. 3a, our model captures the near field hydrodynamics by reproducing the slowing down of the bacterium as it closes in the tracer at $\theta=-\pi/2$ (the bacterium moving towards the center of the spherical tracer), for two typical tracer radii $R_s=1$ and $R_s=100$.  More specifically, our model predictions for both $\bm{F}_{eff}$ and $\bm{v}_{bac}$ show a quantitative agreement with numerical results from Stokeslets method at all ranges of $d$.  Furthermore, our model naturally amended the disastrous overlap problem in the existing two-bead model: in the limiting case where $d\to 0$, the terms in the resistant tensor $\bm{\xi}$ that correspond to the relative motion between the bacterial head and the tracer sphere diverges, leading to an infinitesimal relative motion \cite{RN26}.
\begin{figure}[t]
\includegraphics[width=.48\textwidth]{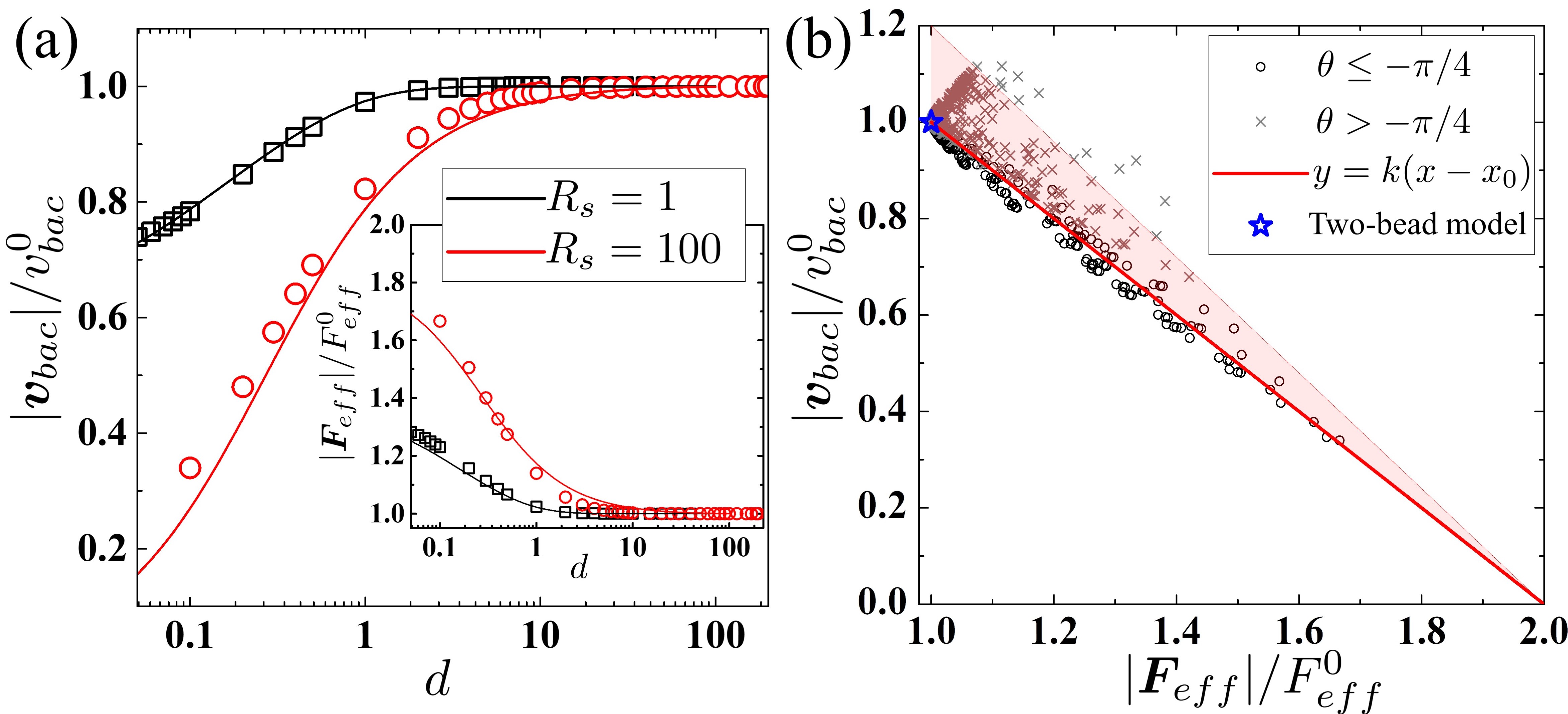}
\caption{Configurational dependent $\bm{v}_{bac}$ and $\bm{F}_{eff}$. (a) Our model predictions based on Eq. 3 (solid lines) and numerical results from the method of Stokeslets (symbols) for $|\bm{v}_{bac} |$ and $|\bm{F}_{eff} |$ (inset) as functions of $d$ at $\theta=-\pi/2$, for $R_s=1$ (black) and $R_s=100$ (red). (b) All numerical data obtained from the method of Stokeslets (symbols) collapse to a master curve as predicted by Eq. 4 (solid line). Shaded area represents $k(x-x_0 )<y<1.2\times k(x-x_0)$. In comparison, existing two-bead model predicts that all data would collapse to a single point (blue star).}
\end{figure}

The proposed idea that the near field hydrodynamic interaction can be modeled entirely through $\bm{\xi}$ is further tested in the study below, for systems more general where analytical form of $\bm{\xi}$ may not be available.  Using the approximation that tail flagella are thin alone (terms in $\bm{\xi}_{tt}$ are constants and much larger than terms in $\bm{\xi}_{st}$), it can be shown that Eq. (3) leads to a generic relation between $\bm{F}_{eff}$ and $\bm{v}_{bac}$ regardless of the shapes of the bacterial head and the tracer (see SI for derivation):
\begin{equation}\label{Eq4}
  \frac{|\bm{v}_{bac}|}{{v}_{bac}^0}=k\times \left( \frac{|\bm{F}_{eff}|}{{F}_{eff}^0}-x_0\right),
\end{equation}
where the slope $k=\frac{{F}_{eff}^0}{{F}_{eff}^0-F_{act}}$ and intercept $x_0=\frac{F_{act}}{{F}_{eff}^0}$  are both intrinsic properties of the bacterium independent of the environment.  As illustrated in Fig. 3b, this generic linear relation is strongly supported by the collapse of all data obtained through the method of Stokeslets onto the predicted straight line, in comparison to the predicted collapse onto a single point $\frac{|\bm{v}_{bac}|}{{v}_{bac}^0} =\frac{|\bm{F}_{eff}|}{{F}_{eff}^0} =1$ by the existing two-bead model.  The agreement between our predicted linear relation and our data shows that model assumptions that capture the qualitative behavior of $\bm{\xi}$ can be sufficient in describing the essence of the near field hydrodynamic interactions, and therefore apply to more general systems with bacterial head (as well as the tracer) of arbitrary non-spherical shapes.

In many non-equilibrium systems, the dynamics of particle pair with small separation is one of the most vital properties in determining the steady state microstructure and the macroscopic collective behaviors \cite{RN27,RN28}.  Therefore, it is critical to investigate the effect due to the near field hydrodynamic interaction on the scattering angle out of the bacterium-tracer pair dynamics (Fig. 4a).  Considering a force-free tracer at the origin and a bacterium at $x=-\infty$ and $y=b$ moving towards $+x$ direction, we studied the dependence of the scattering angle $\psi$ on impact parameters $b/R_s$  and bacterium-tracer size ratio $l/R_s$  (Fig. 4a).  Our proposed model predicts that there exists a critical tracer size in the presence of near field hydrodynamic interactions.  For tracers larger than this critical size, the bacterium can be entrapped by the spherical tracer and swims around the tracer in an orbital motion (Fig. 4b), which can be related to a stable fixed point in the two-dimensional phase plane defined by $d$ and $\theta$ (Fig. 4d-4f).  This entrapment of bacterium has been regularly observed in experiments \cite{RN29,RN30}.  Contrarily, if we replace our tensorial description of the near field hydrodynamics by a repulsive steric interaction, no such entrapment can be reproduced, regardless of the specific form chosen for the steric repulsion (Fig. 4c).  This shows that the resistance tensor $\bm{\xi}$ captures the essence of the near field hydrodynamic interactions and cannot be replaced by any effective steric interactions.  A previous work has also studied the entrapment numerically \cite{RN24}, by simplifying the bacterium as a force dipole and evaluating its near field hydrodynamic interaction with the spherical tracer through the method of images.  However, as the surface distance $d$ becomes very small when the bacterium is entrapped, the dipole treatment of the bacterium becomes an over-simplification insufficient to describe the near field hydrodynamics, since higher order terms in the multipole expansion are also very important \cite{RN31}.  Therefore, unlike our model that predicts a stable fixed point, calculations based on this previous work show that there is only one saddle point in the phase plane (Fig. S10).  And the seeming entrapment observed in this previous work is merely the unphysical consequence of an artificially imposed condition of a minimum surface distance (Fig. S10).
\begin{figure}[t]
\includegraphics[width=0.48\textwidth]{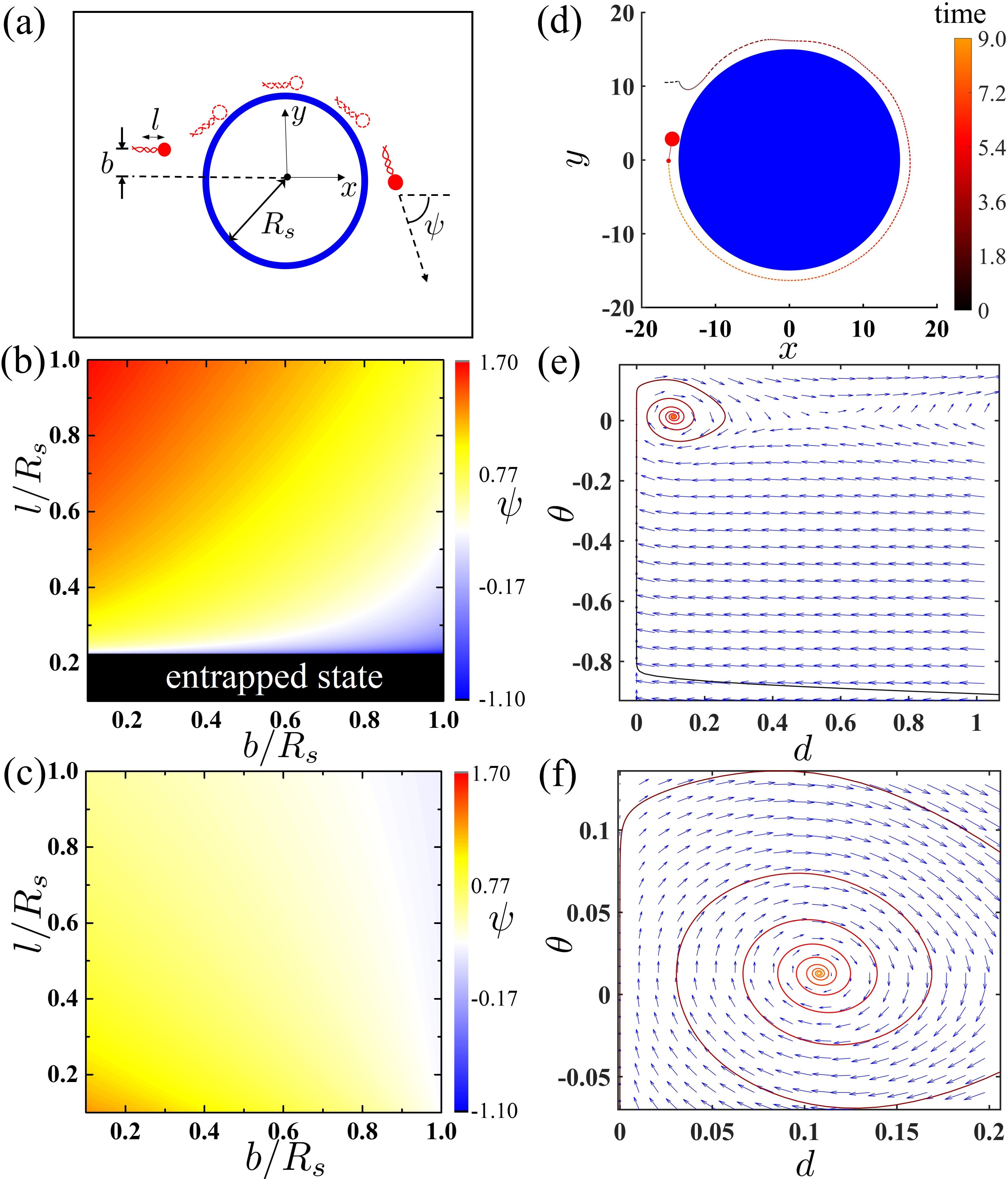}
\caption{Bacterial entrapment. (a) Set-up for the scattering problem, where $\psi$ is positive for counterclockwise rotation. Contour map of $\psi$ is obtained using our proposed model (b), and the existing two-bead model together with a repulsive steric interaction (c), respectively.  During a typical entrapment event, the trajectory of the bacterium is illustrated in real space (d), and in the two-dimensional phase plane defined by $d$ and $\theta$ (e). (f) A closer look of the trajectory in the two-dimensional phase plane, around the stable fixed point.  In (e) and (f), the arrow at each point shows a vector, $(L_A/\sqrt{\dot{d}^2+\dot{\theta}^2})(\dot{d}\hat{\bm{d}}+\dot{\theta}\hat{\bm{\theta}})$, where $\hat{\bm{d}}$ and $\hat{\bm{\theta}}$ are the unit vectors along the $d$ axis and $\theta$ axis, respectively, and the length of the arrow $L_A$ is obtained as a scaled function of $\sqrt{\dot{d}^2+\dot{\theta}^2}$ for illustration purposes (see SI for more information about how $L_A$ is determined for each arrow).}
\end{figure}

In summary, we show that one bacterium interacting with ambient objects can be modeled as two point-forces at fixed locations with a constant propulsion $\bm{F}_{act}$ exerted to the tail bead, while all the local details essential for the description of near field hydrodynamics can be captured through the tensorial coefficient of resistance $\bm{\xi}$ alone.  While the proposed model reduces to the existing two-bead model by keeping only the far field components in $\bm{\xi}$, the artificial overlap problem is naturally avoided by the inclusion of near field components, as the lubrication terms regarding to the relative motion between two objects diverges when the surface distance approaches zero.  We stress that our model is minimal in the sense that it consists of a minimum number of parameters required by the pertaining physical system.  For example, to model an overlap-free bacterial system, our model needs a minimum of only one additional parameter beyond the existing two-bead model---the non-dimensional surface distance, which is the same as the models that use a repulsive steric pair interaction.  More importantly, we would like to point out the essential difference between the proposed model and the existing ones is in the friction nature of the resistance tensor, as opposed to the conservative nature of an effective steric interaction.  The impact of this fundamental difference has been well demonstrated in both passive systems of colloidal suspensions driven out of equilibrium by external shear \cite{RN27} and active suspensions of Janus particles \cite{RN28}, where observations clearly show that the capture of even qualitative behavior of $\bm{\xi}$ through the inclusion of near field lubrication led to very different particle pair dynamics and eventually dramatically different microstructures, in comparison to predictions by models using effective steric interactions.

We thank H. Chate, L. S. Luo and Z. G. Wang for helpful discussions.  This work is supported by NSFC No. 11974038, No. 11672029, No. U1930402, and No. 11904320.  We also acknowledge the computational support from the Beijing Computational Science Research Center.
\bibliography{Draft-2020-PRL}
\end{document}